\documentclass[twocolumn,reprint,amsmath,amssymb,superscriptaddress]{revtex4-1}

\usepackage{graphicx, bm}
\usepackage{fp}
\usepackage{siunitx}
\usepackage{url}
\usepackage{hyperref}
\usepackage{blindtext}
\usepackage{color}
\usepackage[sort&compress]{natbib}
\usepackage{physics}
\usepackage{xprintlen}
\usepackage[colorinlistoftodos,prependcaption,textsize=tiny]{todonotes}
\usepackage{xfrac}
\newcommand{\eqn}[1]{Eq.~\eqref{eqn:#1}}
\newcommand{\fig}[1]{Fig.~\ref{fig:#1}}

\newcommand{\Meff}{\ensuremath{M_\mathrm{eff}}}
\newcommand{\Hk}{\ensuremath{H_\mathrm{k}}}
\newcommand{\Hani}{\ensuremath{H_\mathrm{ani}}}
\newcommand{\Hshape}{\ensuremath{H_\mathrm{shape}}}

\renewcommand{\S}{\ensuremath{S_\mathrm{21}}}
\newcommand{\omegar}{\ensuremath{\omega_\mathrm{res}}}
\newcommand{\dw}{\ensuremath{\Delta \omega}}
\newcommand{\Hres}{\ensuremath{H_\mathrm{res}}}
\newcommand{\Heff}{\ensuremath{H_\mathrm{eff}}}
\newcommand{\dH}{\ensuremath{\Delta H}}
\newcommand{\muB}{\ensuremath{\mu_\mathrm{B}}}

\newcommand{\Tcomp}{\ensuremath{T_\mathrm{comp}}}
\newcommand{\Hc}{\ensuremath{H_\mathrm{c}}}
\renewcommand{\vec}[1]{\mathbf{#1}}
\renewcommand{\hat}[1]{\mathbf{#1}}
\usepackage{color}

\begin{document}

\title{Perpendicular magnetic anisotropy in insulating ferrimagnetic gadolinium iron garnet thin films}

\author{H. Maier-Flaig}
\affiliation{Walther-Mei\ss ner-Institut, Bayerische Akademie der Wissenschaften, Garching, Germany}
\affiliation{Physik-Department, Technische Universit\"{a}t M\"{u}nchen, Garching, Germany}

\author{S. Gepr\"ags}
\affiliation{Walther-Mei\ss ner-Institut, Bayerische Akademie der Wissenschaften, Garching, Germany}

\author{Z. Qiu}
\affiliation{WPI Advanced Institute for Materials Research, Tohoku University, Sendai, Japan}
\affiliation{Spin Quantum Rectification Project, ERATO, Japan Science and Technology Agency, Sendai, Japan}

\author{E. Saitoh}
\affiliation{WPI Advanced Institute for Materials Research, Tohoku University, Sendai, Japan}
\affiliation{Spin Quantum Rectification Project, ERATO, Japan Science and Technology Agency, Sendai, Japan}
\affiliation{Institute for Materials Research, Tohoku University, Sendai, Japan}
\affiliation{PRESTO, Japan Science and Technology Agency, Saitama, Japan}
\affiliation{Advanced Science Research Center, Japan Atomic Energy Agency, Tokai, Japan}

\author{R. Gross}
\affiliation{Walther-Mei\ss ner-Institut, Bayerische Akademie der Wissenschaften, Garching, Germany}
\affiliation{Physik-Department, Technische Universit\"{a}t M\"{u}nchen, Garching, Germany}
\affiliation{Nanosystems Initiative Munich, M\"{u}nchen, Germany}

\author{M. Weiler}
\affiliation{Walther-Mei\ss ner-Institut, Bayerische Akademie der Wissenschaften, Garching, Germany}
\affiliation{Physik-Department, Technische Universit\"{a}t M\"{u}nchen, Garching, Germany}

\author{H. Huebl}
\affiliation{Walther-Mei\ss ner-Institut, Bayerische Akademie der Wissenschaften, Garching, Germany}
\affiliation{Physik-Department, Technische Universit\"{a}t M\"{u}nchen, Garching, Germany}
\affiliation{Nanosystems Initiative Munich,  M\"{u}nchen, Germany}

\author{S. T. B. Goennenwein}
\affiliation{Walther-Mei\ss ner-Institut, Bayerische Akademie der Wissenschaften, Garching, Germany}
\affiliation{Physik-Department, Technische Universit\"{a}t M\"{u}nchen, Garching, Germany}
\affiliation{Nanosystems Initiative Munich, M\"{u}nchen, Germany}
\affiliation{Institut f\"ur Festk\"oper- und Materialphysik, Technische Universit\"at Dresden, Dresden, Germany}
\affiliation{Center for Transport and Devices of Emergent Materials, Technische Universit\"{a}t Dresden, 01062 Dresden}

\date{\today}

\begin{abstract}
We present experimental control of the magnetic anisotropy in a gadolinium iron garnet (GdIG) thin film from in-plane to perpendicular anisotropy by simply changing the sample temperature. 
The magnetic hysteresis loops obtained by SQUID magnetometry measurements unambiguously reveal a change of the magnetically easy axis from out-of-plane to in-plane depending on the sample temperature.
Additionally, we confirm these findings by the use of temperature dependent broadband ferromagnetic resonance spectroscopy (FMR).
In order to determine the effective magnetization, we utilize the intrinsic advantage of FMR spectroscopy which allows to determine the magnetic anisotropy independent of the paramagnetic substrate, while magnetometry determines the combined magnetic moment from film and substrate.
This enables us to quantitatively evaluate the anisotropy and the smooth transition from in-plane to perpendicular magnetic anisotropy.
Furthermore, we derive the temperature dependent $g$-factor and the Gilbert damping of the GdIG thin film. 
\end{abstract}

\maketitle

Controlling the magnetization direction of magnetic systems without the need to switch an external static magnetic field is a challenge that has seen tremendous progress in the past years.
It is of considerable interest for applications as it is a key prerequisite to store information in magnetic media in a fast, reliable and energy efficient way.
Two notable approaches to achieve this in thin magnetic films are switching the magnetization by short laser pulses\cite{Stanciu2007,Lambert2014} and switching the magnetization via spin orbit torques\cite{Garello2014,Miron2011,Brataas2012}. 
For both methods, materials with an easy magnetic anisotropy axis oriented perpendicular to the film plane are of particular interest.
While all-optical switching requires a magnetization component perpendicular to the film plane in order to transfer angular momentum\cite{Lambert2014}, spin orbit torque switching with perpendicularly polarized materials allows fast and reliable operation at low current densities\cite{Garello2014}.
Therefore great efforts have been undertaken to achieve magnetic thin films with perpendicular magnetic anisotropy.\cite{Ikeda2010} However, research has mainly been focused on conducting ferromagnets that are subject to eddy current losses and thus often feature large magnetization damping.
Magnetic garnets are a class of highly tailorable magnetic insulators that have been under investigation and in use in applications for the past six decades.\cite{Calhoun1957,Dionne1971,Adam2002} 
The deposition of garnet thin films using sputtering, pulsed laser deposition or liquid phase epitaxy, and their properties are very well understood.
In particular, doping the parent compound (yttrium iron garnet, YIG) with rare earth elements is a powerful means to tune the static and dynamic magnetic properties of these materials.\cite{Calhoun1957,Belov1961,Roschmann1983}

Here, we study the magnetic properties of a gadolinium iron garnet thin film sample using broadband ferromagnetic resonance (FMR) and SQUID magnetometry.
By changing the temperature, we achieve a transition from the typical in-plane magnetic anisotropy (IPA), dominated by the magnetic shape anisotropy, to a perpendicular magnetic anisotropy (PMA) at about \SI{190}{\kelvin}.
We furthermore report the magnetodynamic properties of GdIG confirming and extending previous results.\cite{Calhoun1957}

\section{Material and sample details}

We investigate a \SI{2.6}{\micro\meter} thick gadolinium iron garnet (Gd$_3$Fe$_5$O$_3$, GdIG) film grown by liquid phase epitaxy (LPE) on a (111)-oriented gadolinium gallium garnet substrate (GGG). 
The sample is identical to the one used in Ref.~\citenum{Maier-Flaig2017-GdIG-coupling} and is described there in detail. 
GdIG is a compensating ferrimagnet composed of two effective magnetic sublattices: 
The magnetic sublattice of the Gd ions and an effective sublattice of the two strongly antiferromagnetically coupled Fe sublattices.
The magnetization of the coupled Fe sublattices shows a weak temperature dependence below room temperature and decreases from approximately \SI{190}{\kilo\ampere\per\meter} at \SI{5}{\kelvin} to \SI{140}{\kilo\ampere\per\meter} at \SI{300}{\kelvin}.\cite{Dionne1971}
The Gd sublattice magnetization follows a Brillouin-like function and decreases drastically from approximately \SI{800}{\kilo\ampere\per\meter} at \SI{5}{\kelvin} to \SI{120}{\kilo\ampere\per\meter} at \SI{300}{\kelvin}.\cite{Dionne1971} 
As the Gd and the net Fe sublattice magnetizations are aligned anti-parallel, the remanent magnetizations cancel each other at the so-called compensation temperature $\Tcomp=\SI{285}{K}$ of the material.\cite{Dionne2009}
Hence, the remanent net magnetization $M$ of GdIG vanishes at $\Tcomp$.

The typical magnetic anisotropies in thin garnet films are the shape anisotropy and the cubic magnetocrystalline anisotropy, but also growth induced anisotropies and magnetoelastic effects due to epitaxial strain have been reported in literature.\cite{Manuilov2009,Manuilov2010}
We find that our experimental data can be understood by taking into account only shape anisotropy and an additional anisotropy field perpendicular to the film plane.
A full determination of the anisotropy contributions is in principle possible with FMR. 
Angle dependent FMR measurements (not shown) indicate an anisotropy of cubic symmetry with the easy axis along the crystal [111] direction in agreement with literature.\cite{Rodrigue1960}
The measurements suggest that the origin of the additional anisotropy field perpendicular to the film plane is the cubic magnetocrystalline anisotropy.
However, the low signal amplitude and the large FMR linewidth towards $\Tcomp$ in combination with a small misalignment of the sample, render a complete, temperature dependent anisotropy analysis impossible.
In the following, we therefore focus only on shape anisotropy and the additional out-of-plane anisotropy field.

\section{SQUID magnetometry}
\begin{figure}
\includegraphics[width=0.49\textwidth]{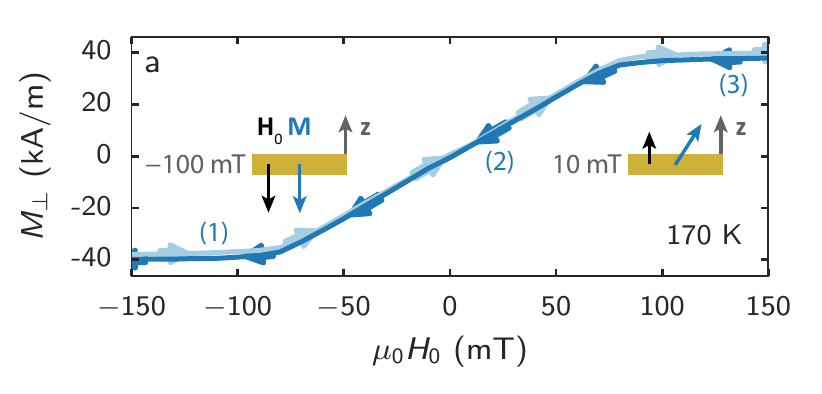}
\includegraphics[width=0.49\textwidth]{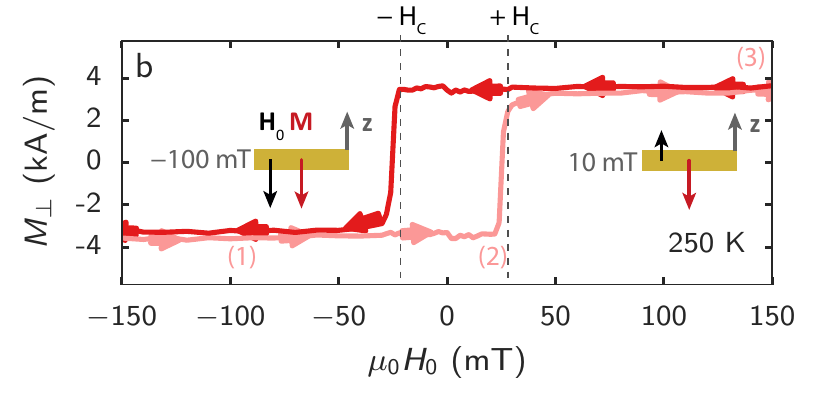}
\caption{
  Out-of-plane magnetization component $M_\perp$ measured by SQUID magnetometry.
  For different temperatures, magnetically hard (170K, (a)) and easy (250K, (b)) axis loops are observed. The arrows on the data indicate the sweep direction of $H_0$.
  The insets schematically show the magnetization direction $\vec{M}$ and $\vec{H}_0 = H_0 \hat z$ with the film normal $\hat z$ at the indicated values of $H_0$.
}
\label{fig:squid}
\end{figure}
SQUID magnetometry measures the projection of the magnetic moment of a sample on the applied magnetic field direction. 
For thin magnetic films, however, the background signal from the comparatively thick substrate can be on the order of or even exceed the magnetic moment $m$ of the thin film and hereby impede the quantitative determination of $m$.
Our \SI{2.6}{\micro\meter} thick GdIG film is grown on a \SI{500}{\micro\meter} thick GGG substrate warranting a careful subtraction of the paramagnetic background signal of the substrate.
In our experiments, $H_0$ is applied perpendicular to the film plane and thus, the projection of the net magnetization $\vec{M} = \vec{m}/V$ to the out-of-plane axis is recorded as $M_\perp$.
\fig{squid} shows $M_\perp$ of the GdIG film as function of the externally applied magnetic field $H_0$.
In the investigated small region of $H_0$, the magnetization of the paramagnetic substrate can be approximated by a linear background that has been subtracted from the data.
The two magnetic hysteresis loops shown in \fig{squid} are typical for low temperatures ($T\lesssim \SI{170}{\kelvin}$) and for temperatures close to $\Tcomp$.
The hysteresis loops unambiguously evidence hard and easy axis behavior, respectively.
Towards low temperatures ($T=\SI{170}{\kelvin}$, \fig{squid}\,(a)) the net magnetization $M=|\vec{M}|$ increases and hence, the anisotropy energy associated with the demagnetization field $\Hshape = -M_\perp$%
\footnote{
  We use the demagnetization factors of a infinite thin film: $N_\mathrm{x,y,z} = (0,0,1)$.
}
dominates and forces the magnetization to stay in-plane. 
At these low temperatures, the anisotropy field perpendicular to the film plane, $\Hk$, caused by the additional anisotropy contribution has a constant, comparatively small magnitude. 
We therefore observe a hard axis loop in the out-of-plane direction: 
Upon increasing $H_0$ from \SI{-150}{\milli\tesla} to \SI[retain-explicit-plus]{+150}{\milli\tesla}, $\vec{M}$ continuously rotates from the out-of-plane (oop) direction to the in-plane (ip) direction and back to the oop direction again.
The same continuous rotation happens for the opposite sweep direction of $H_0$ with very little hysteresis.
For temperatures close to $\Tcomp$ ($T=\SI{250}{\kelvin}$, \fig{squid}\,(b)), $\Hshape$ becomes negligible due to the decreasing $M$ while $\Hk$ increases as shown below.
Hence, the out-of-plane direction becomes the magnetically easy axis and, in turn, an easy-axis hysteresis loop is observed: 
After applying a large negative $H_0$ [(1) in \fig{squid}\,(a)] $M$ and $H_0$ are first parallel. Sweeping to a positive $H_0$, $M$ first stays parallel to the film normal and thus $M_\perp$ remains constant [(2) in \fig{squid}\,(a)] until it suddenly flips to being aligned anti-parallel to the film normal at $H_0 > +\Hc$ [(3) in \fig{squid}\,(a)].
These loops clearly demonstrate that the nature of the anisotropy changes from IPA to PMA on varying temperature.

\section{Broadband ferromagnetic resonance}
\begin{figure*}
\includegraphics[width=0.95\textwidth]{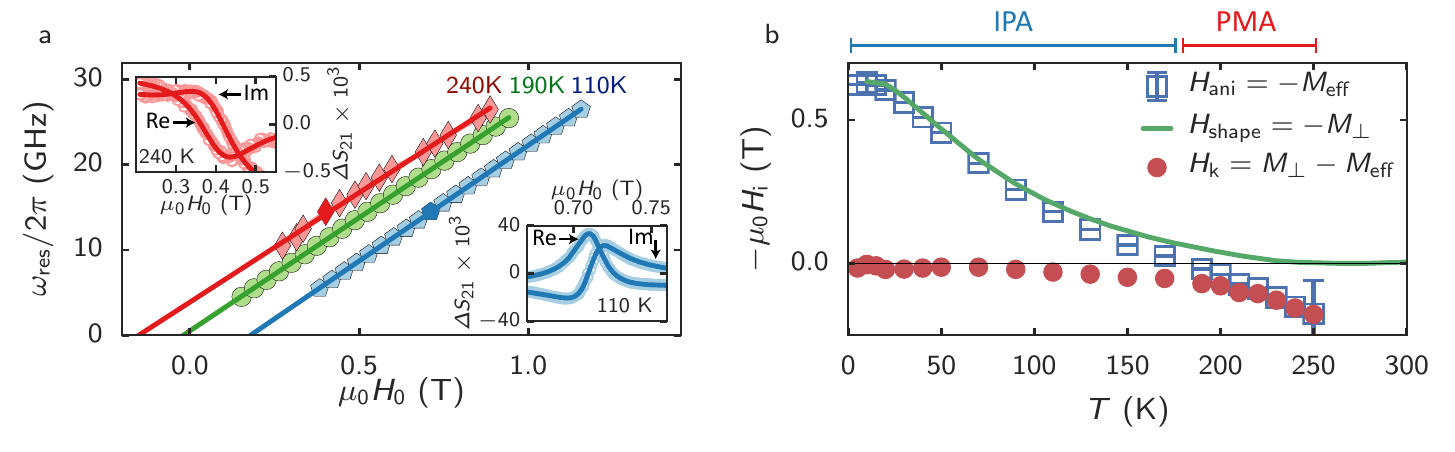}
\caption{
  Broadband FMR spectroscopy data reveiling a smooth transition from in-plane to perpendicular anisotropy.
  \textbf{(a)} FMR resonance frequency plotted against $H_0$ taken for three different temperatures (symbols) and fit to \eqn{oop_dispersion} (solid lines). 
  For an IPA, a positive effective magnetization $\Meff$ (positive $x$-axis intercept) is extracted, whereas $\Meff$ is negative for a PMA.
  \textbf{(inset)} Exemplary resonance spectra (symbols) at \SI{14.5}{\giga\hertz} recorded at \SI{110}{\kelvin} and \SI{240}{\kelvin} as well as the fits to \eqn{s21} used to determine $\omegar$ (solid lines). 
  A complex offset $\S^0$ has been subtracted for visual clarity, plotted is $\Delta \S = \S-\S^0$.
  \textbf{(b)} Anisotropy field $\Hani=-\Meff$ as a function of temperature (open squares). Prediction for shape anisotropy $\Hshape$ based on SQUID magnetometry data (solid line) from Ref.~\citenum{Geprags2016}. The additional perpendicular anisotropy field $\Hk= M_\perp - \Meff$ (red dots) increases to approximately \SI{0.18}{\tesla} at \SI{250}{\kelvin} where its value is essentially identical to $\Hani$ due to the vanishing $M_\perp$.
  }
\label{fig:fmr}
\end{figure*}

In order to quantify the transition from in-plane to perpendicular anisotropy found in the SQUID magnetometry data, broadband FMR is performed as a function of temperature with the external magnetic field $H_0$ applied along the film normal.\footnote{
  The alignment of the sample is confirmed at low temperatures by performing rotations of the magnetic field direction at fixed magnetic field magnitude while recording the frequency of resonance $\omegar$.
  As the shape anisotropy dominates at low temperatures, $\omegar$ goes through an easy-to-identify minimum when the sample is aligned oop.
  }
For this, $H_0$ is swept while the complex microwave transmission $\S$ of a coplanar waveguide loaded with the sample is recorded at various fixed frequencies between \SI{10}{\giga\hertz} and \SI{25}{\giga\hertz}.
We perform fits of $\S$ to\cite{Maier-Flaig2017-DD}
\begin{equation}\label{eqn:s21}
  \S\left(H_0\right)|_\omega =\
   -i\,Z \chi\!\left(H_0\right) + A + B\cdot H_0
\end{equation}
with the complex parameters $A$ and $B$ accounting for a linear field-dependent background signal of \S, the complex FMR amplitude $Z$, and the Polder susceptibility\cite{Shaw2013a,Nembach2011}
\begin{equation}
  \label{polder}
  \chi\left(H_0\right)  
  = \frac{\Meff \left(H - \Meff \right)}
         {\left(H - \Meff \right)^2 - \Heff^2 + i\frac{\dH}{2}\left(H - \Meff \right)}.
\end{equation}
Here, $\gamma$ is the gyromagnetic ratio, $\Heff = \omega/(\gamma \mu_0)$, and $\omega$ is the microwave frequency and the effective magnetization $\Meff=\Hres - \omegar/(\gamma \mu_0)$.
From the fit, the resonance field $H_\mathrm{res}$ and the full width at half-maximum (FWHM) linewidth $\dH$ is extracted.
Exemplary data for $\S$ (data points) and the fits to \eqn{s21} (solid lines) at two distinct temperatures are shown in the two insets of \fig{fmr}\,(a). 
We obtain excellent agreement of the fits and the data.
The insets furthermore show that the signal amplitude is significantly smaller for $T=\SI{240}{\kelvin}$ than for $\SI{110}{\kelvin}$. 
This is expected as the signal amplitude is proportional to the net magnetization $M$ of the sample which decreases considerably with increasing temperature (cf. \fig{fmr}\,(b)).
At the same time, the linewidth drastically increases as discussed in the following section.
These two aspects prevent a reliable analysis of the FMR signal in the temperature region $\SI{250}{\kelvin} < T < \SI{300}{\kelvin}$ (i.e. around the compensation temperature). 
Therefore we do not report data in this temperature region. 
Nevertheless, FMR is ideally suited to investigate the magnetic properties of the GdIG film selectively, i.e. independent of the substrate, for temperatures below $\Tcomp$.

As all measurements are performed in the high field limit of FMR, the dispersions shown in \fig{fmr}\,(a) are linear and we can use the Kittel equation
\begin{equation}\label{eqn:oop_dispersion}
  \omegar = \gamma \mu_0 \left( \Hres - \Meff \right)
\end{equation}
to extract $\gamma$ and $\Meff$.
It is customary to describe the magnetic anisotropy using $\Meff$ which can be related to an anisotropy field $\Hani$ along $+\hat z$ as $\Meff=-\Hani=M_\perp - \Hk$ for positive $H_0$.
Here, $\Hani$ is given by the demagnetization field $\Hshape = -M_\perp$ (along $-\hat z$) and the anisotropy field $\Hk$ of the additional perpendicular anisotropy (along $+\hat z$).
Evidently, $\Meff$ can be determined by linearly extrapolating the data to $\omegar=0$.
The FMR dispersion and the fit to \eqn{oop_dispersion} (solid lines) are shown for three selected temperatures in \fig{fmr}\,(a).
At \SI{110}{\kelvin} (blue curve) $\Meff$ is positive.
Therefore, $M > \Hk$ indicating that shape anisotropy dominates, and the film plane is a magnetically easy plane while the oop direction is a magnetically \textit{hard} axis.
At \SI{240}{\kelvin} (red curve) $\Meff$ is negative and hence, the oop direction is a magnetically \textit{easy} axis. 
Figure~ \ref{fig:fmr}\,(b) shows the extracted $\Meff(T)$. At \SI{190}{\kelvin}, $\Meff$ changes sign. Above this temperature (marked in red), the oop axis is magnetically easy (PMA) and below this temperature (marked in blue), the oop axis is magnetically hard (IPA).
The knowledge of $M_\perp\!\left(T\right)$ obtained from SQUID measurements allows to separate the additional anisotropy field $\Hk$ from $\Meff$ (red dots in \fig{fmr}~(b)).
$\Hk=M_\perp-\Meff$ increases considerably for temperatures close to $\Tcomp$ while at the same time the contribution of the shape anisotropy, $\Hshape = - M_\perp$ trends to zero.
For $T \gtrapprox \SI{180}{\kelvin}$, $\Hk$ exceeds $\Hshape$ which is indicated by the sign change of $\Meff$. 
Above this temperature, we thus observe PMA.
We use the magnetization $M$ determined using SQUID magnetometry from Ref.~\citenum{Geprags2016} normalized to the here recorded $\Meff$ at $\SI{10}{\kelvin}$ in order to quantify $\Hk$.
The maximal value $\mu_0 \Hk = \SI{0.18}{\tesla}$ is obtained at \SI{250}{\kelvin} which is the highest measured temperature due to the decreasing signal-to-noise ratio towards $\Tcomp$.

\begin{figure}
\includegraphics[width=0.49\textwidth]{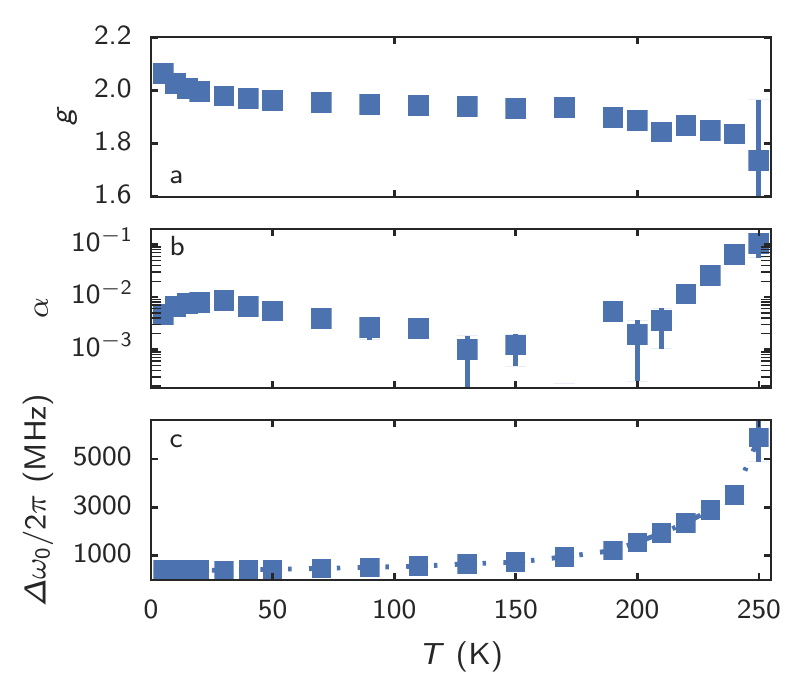}
\caption{
  Key parameters characterizing the magnetization dynamics of GdIG as a function of $T$:
  \textbf{(a)} $g$-factor $g=\gamma \hbar/\mu_\mathrm{B}$,
  \textbf{(b)} Gilbert damping constant $\alpha$ and 
  \textbf{(c)} inhomogeneous linewidth $\dw_0/(2\pi)$
  }
\label{fig:g_damping}
\end{figure}

We can furthermore extract the $g$-factor and damping parameters from FMR. The evolution of the $g$-factor $g = \gamma \frac{\hbar}{\muB}$ with temperature is shown in \fig{g_damping}\,(a).
We observe a substantial decrease of $g$ towards $\Tcomp$. This is consistent with reports in literature for bulk GIG and can be explained considering that the $g$-factors of Gd and Fe ions are slightly different such that the angular momentum compensation temperature is larger than the magnetization compensation temperature.\cite{Wangsness1956}
The linewidth $\dw = \gamma \dH$ can be separated into a inhomogeneous contribution $\dw_0 = \dw(H_0 = 0)$ and a damping contribution varying linear with frequency with the slope $\alpha$: 
\begin{equation}\label{eqn:gilbert}
  \dw = 2 \alpha \cdot \omegar + \dw_0.
\end{equation}
Close to $\Tcomp=\SI{285}{K}$, the dominant contribution to the linewidth is $\dw_0$ which increases by more than an order of magnitude from \SI{390}{\mega\hertz} at \SI{10}{K} to \SI{6350}{\mega\hertz} at \SI{250}{K} [\fig{g_damping}\,(c)].
This temperature dependence of the linewidth has been described theoretically by Clogston et al.\cite{Clogston1958,Geschwind1957} in terms of a dipole narrowing of the inhomogeneous broadening and was reported experimentally before\cite{Rodrigue1960,Calhoun1957}.
As opposed to these single frequency experiments, our broadband experiments allow to separate inhomogeneous and intrinsic damping contributions to the linewidth.
We find that in addition to the inhomogeneous broadening of the line, also the Gilbert-like (linearly frequency dependent) contribution to the linewidth changes significantly:
Upon approaching $\Tcomp$ [\fig{g_damping}\,(b)], the Gilbert damping parameter $\alpha$ increases by an order of magnitude. 
Note, however, that due to the large linewidth and the small magnetic moment of the film, the determination of $\alpha$ has a relatively large uncertainty.%
\footnote{
  For the given signal-to-noise ratio and the large linewidth, $\alpha$ and $\dw_0$ are correlated to a non-negligible degree with a correlation coefficient of $C(\mathrm{intercept}, \mathrm{slope}) = -0.967$.
  }
A more reliable determination of the temperature evolution of $\alpha$ using a single crystal GdIG sample that gives access to the intrinsic bulk damping parameters remains an important task.

\section{Conclusions}
We investigate the temperature evolution of the magnetic anisotropy of a GdIG thin film using SQUID magnetometry as well as broadband ferromagnetic resonance spectroscopy.
At temperatures far away from the compensation temperature $\Tcomp$, the SQUID magnetometry reveals hard axis hysteresis loops in the out-of-plane direction due to shape anisotropy dominating the magnetic configuration. 
In contrast, at temperatures close to the compensation point, we observe easy axis hysteresis loops. 
Broadband ferromagnetic resonance spectroscopy reveals a sign change of the effective magnetization (the magnetic anisotropy field) which is in line with the magnetometry measurements and allows a quantitative analysis of the anisotropy fields.
We explain the qualitative anisotropy modifications as a function of temperature by the fact that the magnetic shape anisotropy contribution is reduced considerably close to $\Tcomp$ due to the reduced net magnetization, while the additional perpendicular anisotropy field increases considerably.
We conclude that by changing the temperature the nature of the magnetic anisotropy can be changed from an in-plane magnetic anisotropy to a perpendicular magnetic anisotropy.
This perpendicular anisotropy close to $\Tcomp$ in combination with the small magnetization of the material may enable optical switching experiments in insulating ferromagnetic garnet materials.
Furthermore, we analyze the temperature dependence of the FMR linewidth and the $g$-factor of the GdIG thin film where we find values compatible with bulk GdIG\cite{Geschwind1957,Calhoun1957}. 
The linewidth can be separated into a Gilbert-like and an inhomogeneous contribution.
We show that in addition to the previously reported increase of the inhomogeneous broadening, also the Gilbert-like damping increases significantly when approaching $\Tcomp$

\section{Acknowledgments}
We gratefully acknowledge funding via the priority program Spin Caloric Transport (spinCAT), (Projects GO 944/4 and GR 1132/18), the priority program SPP 1601 (HU 1896/2-1) and the collaborative research center SFB 631 of the Deutsche Forschungsgemeinschaft.

\section{Bibliography}

\bibliography{references}

\end{document}